\documentclass[pss]{wiley2sp} 
\usepackage{amsmath}

\begin{document}

\title{Ferromagnetic quantum phase transition in Sr$_{1-x}$Ca$_x$RuO$_3$ thin films}

\titlerunning{FM QPT in Sr$_{1-x}$Ca$_x$RuO$_3$ thin films}

\author{M. Schneider, V. Moshnyaga, P. Gegenwart{\Ast}}

\authorrunning{M. Schneider et al.}

\mail{e-mail
  \textsf{pgegenw@gwdg.de}, Phone:
  +49-551-397607, Fax: +49-551-3919546}

\institute{I. Physikalisches Institut, Georg-August-Universit\"{a}t
G\"{oe}ttingen, Friedrich-Hund-Platz 1, 37077  G\"{oe}ttingen,
Germany}


\received{XXXX, revised XXXX, accepted XXXX} 
\published{XXXX} 


\abstract{%
%
%
%
\abstcol{%
The ferromagnetic quantum phase transition in the perovskite
ruthenate Sr$_{1-x}$Ca$_x$RuO$_3$ is studied by low-temperature
magnetization and electrical resistivity measurements on thin films.
The films were grown epitaxially on SrTiO$_3$ substrates using
metalorganic aerosol deposition and characterized by X-ray
diffraction and room temperature scanning tunneling microscopy. High
residual resistivity ratios of 29 and 16 for $x=0$ and $x=1$,
respectively, prove the high quality of the investigated samples.
  }{%
  We observe a continuous suppression of the ferromagnetic
Curie temperature from $T_C=160$~K at $x=0$ towards $T_C\rightarrow
0$ at $x_c\approx 0.8$. The analysis of the electrical resistivity
between 2 and 10~K reveals $T^2$ and $T^{3/2}$ behavior at $x\leq
0.6$ and $x\geq 0.7$, respectively. For undoped CaRuO$_3$, the
measurement has been extended down to 60~mK, revealing a crossover
to $T^2$ behavior around 2~K, which suggests a Fermi-liquid ground
state in this system.}}

%
%

\maketitle   

\paragraph{Introduction}

Quantum phase transitions (QPTs) in weak itinerant ferromagnets have
recently attracted much interest due to the discovery of exciting
low-temperature states such as unconventional superconductivity
\cite{Saxena}, partial magnetic order \cite{Pfleiderer 04} or
non-Fermi liquid (NFL) phases \cite{Lonzarich}. The system
Sr$_{1-x}$Ca$_x$RuO$_3$ represents a rare example of an oxide NFL
metal that displays an itinerant electron ferromagnetic (FM) QPT
\cite{Cao,Yoshimura,Khalifah}. Although the general trend of a
continuous suppression of ferromagnetism with increasing Ca
concentration is well established, details close to the QPT differ
substantially in previous reports. Early studies on sintered
polycrystals have suggested a FM quantum critical point at $x=0.7$
\cite{Yoshimura}. Subsequent Muon-spin rotation measurements on the
same crystals have proven that the magnetically ordered volume
fraction at low-$T$ decreases from 100\% upon increasing Ca content
$x$ well {\it before} $T_C\rightarrow 0$~\cite{Uemura}, indicating
magnetic phase separation. Most interestingly, similar behavior has
also been found near the pressure-driven first-order QPT in MnSi
\cite{Uemura}. On the other hand, magnetization measurements on flux
grown single crystals have shown FM order beyond $x=0.8$ with a
smooth crossover to spin-glass-like behavior at larger $x$,
extending towards $x=1$~\cite{Cao}. Electrical resistivity
measurements on thin films, prepared by pulsed-laser deposition have
revealed an anomalous temperature dependence
$\rho(T)=\rho_0+A'T^{3/2}$ in a wide concentration range $0.8\leq
x\leq 1$ at temperatures between 2 and 10~K, which resembles the NFL
phase observed in MnSi under hydrostatic pressure. However, these
thin films have a poor residual resistivity ratio, even for the
undoped systems SrRuO$_3$ and CaRuO$_3$ and the experiments should
be extended towards lower temperatures. In this paper, we report
magnetization and electrical resistivity measurements on epitaxial
Sr$_{1-x}$Ca$_x$RuO$_3$ thin films that have been grown using
metal-organic aerosol deposition (MAD) \cite{ICM paper}.

\paragraph{Experimental}

Thin films of Sr$_{1-x}$Ca$_x$RuO$_3$, where grown by MAD on MgO and
SrTiO$_3$ substrates \cite{ICM paper}. In the former case, the
larger lattice mismatch leads to a polycrystalline growth of
islands, resulting in a substantial film roughness. By contrast, in
the latter case epitaxial growth was obtained with very small values
of the roughness detected by room temperature scanning tunneling
microscopy. In this paper, we focus only on films grown on SrTiO$_3$
with in-plane dimensions of $10\times 5$~mm. Details of their
preparation and structural characterization by X-ray diffraction and
room-temperature scanning tunneling microscopy are reported in
\cite{ICM paper}. The thickness of each film has been determined by
small-angle X-ray scattering and varies between 40 and 50~nm.
Magnetization and electrical resistivity measurements at
temperatures between 2 and 300~K where preformed using Quantum
Design MPMS and PPMS systems, respectively. The electrical
resistivity between 60~mK and 10~K has been determined in an
adiabatic demagnetization cryostat on a microstructured film.
Electrical contacts were made using silver paste.

%

\paragraph{Magnetization}

\begin{figure}
\begin{center}
\includegraphics[width=\linewidth,keepaspectratio]{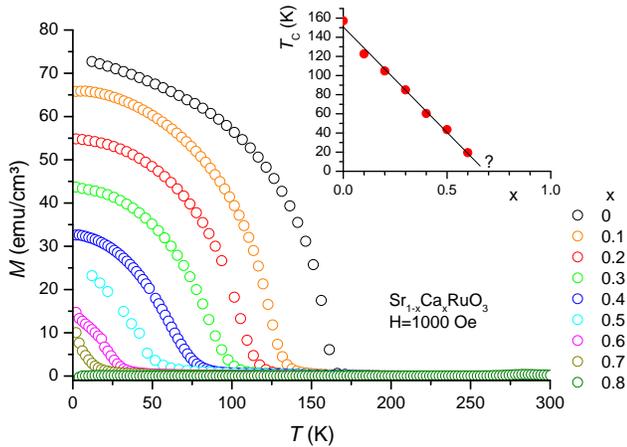}
\end{center}
\caption{\label{label}Temperature dependence of the magnetization of
Sr$_{1-x}$Ca$_x$RuO$_3$ thin films. Each data set has been obtained
at $H=1000$~Oe  after subtraction of the magnetization of a
SrTiO$_3$ substrate. The inset displays the Curie-Temperature $T_C$
vs $x$, derived from the inflection points in $M(T)$.}
\end{figure}

To determine the magnetization of the Sr$_{1-x}$Ca$_x$RuO$_3$ thin
films, the background contribution of the SrTiO$_3$ substrate must
be subtracted. A measurement on a plain substrate has revealed a
diamagnetic magnetization with an additional Curie-Weiss
contribution at low temperatures resulting from a small amount of
magnetic impurities. While the substrate contribution to the total
magnetization is small for FM samples below their Curie temperature,
it dominates in the paramagnetic regime in particular for $x\leq
0.7$ at low temperatures. Since our analysis assumes similar
substrate contributions for all samples, which, however, may be
incorrect for substrates with differing amount of magnetic
impurities, a substantial error arises at low temperatures and large
Ca concentrations. Thus, the exact position of the QPT cannot be
determined from these experiments. As estimate for $T_C$, we have
analyzed the positions of the inflection points of $M(T)$ measured at $H=1000$~Oe.\\
As displayed in Fig.~1, the so-derived $T_C$ data indicate a clear
and systematic evolution from a FM in ground state at $x=1$ towards
a paramagnetic ground state at $x=0.8$, in good agreement with
\cite{Khalifah,Uemura}. For a closer analysis of the magnetic
properties close to the FM QPT it would be required to measure for
each film the substrate contribution {\it before} the synthesis and
to explore the exact positions of $T_C(x)$ by Arrott-plot analysis.

\paragraph{Electrical resistivity}

\begin{table}[b]
  \caption{Parameters derived from fits of the
electrical resistivity of Sr$_{1-x}$Ca$_x$RuO$_3$ thin films at
temperatures between 2 and 10~K according to $R(T)=R_0+AT^n$. RRR
denotes residual resistivity ratio $R_{300K}/R_0$.}
  \begin{tabular}[htbp]{@{}lllll@{}}
    \hline
    Sample & $x$ & $A/R_{300K}$ ($10^{-5}$K$^{-n}$) & $n$ & RRR\\
    \hline
    T449 & 0 & ($7.86\pm 0.14$) & $2.09\pm 0.02$ & 28.9\\
    T648 & 0.1 & ($4.66\pm 0.53$) & $2.11\pm 0.04$ & 4.2\\
    T643 & 0.2 & ($6.30\pm 0.06$) & $2.05\pm 0.01$ & 3.7\\
    T642 & 0.3 & ($3.18\pm 0.80$) & $2.13\pm 0.05$ & 3.6\\
    T644 & 0.4 & ($9.51\pm 0.21$) & $2.01\pm 0.02$ & 3.6\\
    T414 & 0.5 & ($8.26\pm 2.1$) & $2.03\pm 0.09$ & 2.9\\
    T499 & 0.6 & ($25.8\pm 2.4$) & $1.89\pm 0.02$ & 3.3\\
    T498 & 0.7 & ($62.0\pm 8.8$) & $1.57\pm 0.07$ & 3.2\\
    T496 & 0.8 & ($73.9\pm 8.9$) & $1.48\pm 0.05$ & 4.2\\
    T495 & 0.9 & ($69.7\pm 9.5$) & $1.53\pm 0.02$ & 6.1\\
    T440 & 1.0 & ($79.4\pm 2.0$) & $1.51\pm 0.03$ & 15.7\\
    \hline
  \end{tabular}
  \label{onecolumntable}
\end{table}

\begin{figure}
\begin{center}
\includegraphics[width=\linewidth,keepaspectratio]{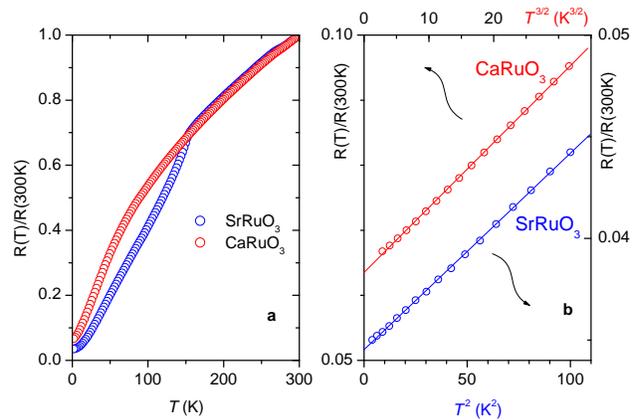}
\end{center}
\caption{\label{label}(a): Temperature dependence of the electrical
resistivity of SrRuO$_3$ (circles) and CaRuO$_3$ (squares) thin
films. (b): Same data below 10~K plotted versus $T^2$ and $T^{3/2}$
for SrRuO$_3$ and CaRuO$_3$, respectively.}
\end{figure}

We now turn to the temperature dependence of the electrical
resistivity $R(T)$ of the Sr$_{1-x}$Ca$_x$RuO$_3$ thin films.
Fig.~2a displays normalized data at temperatures between 2 and 300~K
for $x=0$ and $x=1$. Similar measurements have been performed for
all concentrations listed in Table~1. At low Ca concentrations,
$x\leq 0.5$, sharp decreases in $R(T)$ are observed upon cooling
through the respective Curie temperatures. For larger $x$, the
signature becomes very weak and disappears. The residual resistivity
ratios of 29 and 16 for $x=0$ and $x=1$, respectively, are much
better compared to those reported in \cite{Khalifah,Klein} and
indicate the good quality of the films. The lowest RRR value, is
found for $x=0.5$ (cf. Table~1), as expected from the maximum
disorder due to the random Sr- and Ca substitution. The low-$T$ data
between 2 and 10~K have been fitted by a power-law temperature
dependence, whose parameters are given in Table~1. Whereas a $T^2$
dependence, as expected for Fermi liquids, is found at $x<0.7$,
clear deviation is observed at larger Ca content. As shown in
Fig.~2b for $x=1$, the resistivity between 2 and 10~K could be
described by a $T^{3/2}$ dependence, similar as found previously for
thin films \cite{Khalifah,Klein} as well as single crystals in the
same temperature range \cite{Cao,Kikugawa}. The extension of the
$x=1$ data towards lower $T$ is discussed later.

\begin{figure}
\begin{center}
\includegraphics[width=\linewidth,keepaspectratio]{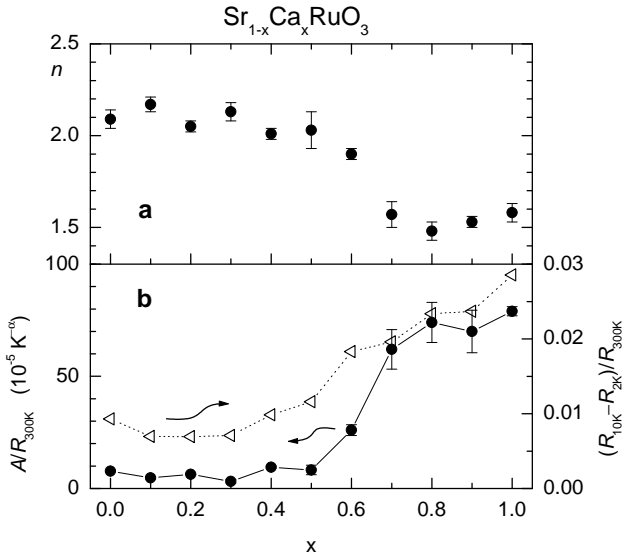}
\end{center}
\caption{\label{label}(a): Temperature exponent $n$ derived from the
fitting described in Table~1, versus Ca concentration in
Sr$_{1-x}$Ca$_x$RuO$_3$. (b): Corresponding coefficient $A/R_{300K}$
(left axis) and resistance increase between 2 and 10~K, normalized
to the room temperature resistance (right axis).}
\end{figure}

Fig.~3 summarizes the concentration dependence of the fit parameters
given in Table~1. Since it is difficult to compare values of the
coefficient $A$ obtained for varying exponents $n$, we also show the
difference of the resistance between 2 and 10~K, normalized to the
room temperature resistance. This latter property could be taken as
a measure for the low-$T$ scattering due to magnetic fluctuations in
this system. A clear increase is found with increasing
Ca-concentration suggesting enhanced magnetic fluctuations near the
disappearance of long-range FM order. This is also compatible with
an enhanced specific heat coefficient ($\gamma\approx
80$~mJ/(K$^2$mol-Ru) in CaRuO$_3$~\cite{Kikugawa}).

\begin{figure}
\begin{center}
\includegraphics[width=\linewidth,keepaspectratio]{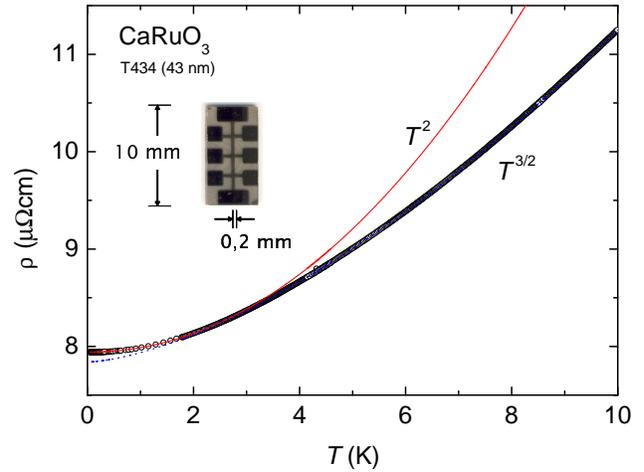}
\end{center}
\caption{\label{label}Low-temperature electrical resistivity of
CaRuO$_3$ thin film number T434. The red solid and blue dotted lines
display $T^2$ and $T^{3/2}$ behaviors, respectively. A photograph of
the microstructured film with thickness of 43 nm is shown in the
inset.}
\end{figure}

In order to study a possible NFL ground state in CaRuO$_3$, the
resistivity measurement has been extended down to 60~mK using a
microstructured film shown in Fig.~4. Remarkably, we observe a
crossover from $T^{3/2}$ to $T^2$ behavior below 2~K. This is in
accordance with recent specific heat measurements on a high-quality
single crystal, which show a constant specific heat coefficient at
low temperatures~\cite{Kikugawa}. Similar low-$T$ measurements on
thin films with $0.8\leq x\leq 1$ are underway to determine the
exact region in the $T-x$ phase diagram where the resistivity
displays NFL behavior.

\paragraph{Conclusion}

The FM QPT in Sr$_{1-x}$Ca$_x$RuO$_3$ thin films has been studied
using magnetization and electrical resistivity measurements. A clear
and systematic evolution from $T_C=160$~K at $x=0$ towards
$T_C\rightarrow 0$ at $x\approx 0.7$ is observed. The analysis of
the temperature dependence of the electrical resistivity between 2
and 10~K suggests a broad concentration range in which non-Fermi
liquid like behavior ($\Delta\rho\propto T^{3/2}$) is observed.
However, the extension of the data for undoped CaRuO$_3$ down to
60~mK has revealed a crossover to Fermi liquid behavior
$\Delta\rho\propto T^{2}$ at $T\leq 2$~K.

\begin{acknowledgement}
We thank C. Stingl and K. Winzer for their help with the low-$T$
electrical resistivity measurement. Work supported by DFG through
SFB 602.
\end{acknowledgement}

%
%

\end{document}